\documentclass[lineno]{JFM-FLM_Au}

\usepackage{hyperref}
\usepackage{cleveref}

\lefttitle{Mark Warnecke}
\righttitle{Journal of Fluid Mechanics}

\title{Multipoint Statistical Turbulent Dynamics from Hopf Equation Closures}

\author{Mark Warnecke\aff{1}}

\affiliation{\aff{1}Mechanical \& Aerospace Engineering, University of California Irvine.}

\corresau{Mark Warnecke, \email{mwarneck@uci.edu}}

\begin{document}
\maketitle

\begin{abstract}
Obtaining accurate  multipoint statistics of turbulence is computationally very expensive and therefore these statistics have remained largely unexplored from a theoretical standpoint. In this paper, (i) a first-principles-based closure of the $n$th-order structure function governing equation proposed by \citet{SreenivasanYakhot2021moments} is generalized to a closure of the velocity increment Hopf equation itself. Then (ii) the closure is further generalized to the $N$-point Hopf equation. Finally, (iii) an example of the method is provided to analytically determine the $3$-point structure function transition between the known $2$-point structure function and the $3$-point fusion rules from the closed $(N=3)$-point velocity increment Hopf equation. The analytical solution takes the form of a Batchelor interpolation and shows promising agreement with preliminary DNS data for the cases examined. Since the $N$-point velocity increment Hopf equation is closed, its solution can be numerically approximated. It is expected that similar methods, applied here to obtain the $2$-point structure function and $3$-point structure function transition, can be used to obtain further analytical predictions of various multipoint quantities to deepen our understanding of turbulence.
\end{abstract}

\begin{keywords}
Authors should not enter keywords on the manuscript, as these must be chosen by the author during the online submission process and will then be added during the typesetting process (see \href{https://www.cambridge.org/core/journals/journal-of-fluid-mechanics/information/list-of-keywords}{Keyword PDF} for the full list).  Other classifications will be added at the same time.
\end{keywords}


\section{Introduction}

Turbulence is characterized by nonlinear, chaotic, and multiscale continuous fluid motion governed by the Navier-Stokes equations. The degree of turbulence is captured by the Reynolds number, $Re = UL/\nu$, where $U$ is a characteristic velocity scale at the scale $L$ and $\nu$ is the kinematic viscosity of the fluid.
In the very high Reynolds number limit, turbulence contains a wide range of interacting and interdependent dynamic scales; this fact makes analytical progress extremely difficult and numerical simulation computationally demanding. 

The nonlinear chaotic nature of the Navier-Stokes equations and the practical limits of real-world measurements mean that turbulence is practically unpredictable from a deterministic point of view. Any tiny measurement error in the velocity field will rapidly grow and propagate, leading to a large mismatch between a computational prediction and a later measurement of the flow field. Due to this fact, and the engineering pressure to concentrate on averaged flow field quantities, the only experimentally reproducible, measurable, and (in principle) predictable quantities are statistical 
\citep{monin1971vol1and2, hopf1952jrma}. In this formalism, the velocity field $\overline{u}(\overline{x})$ is treated as a random variable that is fully characterized by its probability measure. $N$-point statistics are defined as statistical measures captured by the probability measures (PDFs) of the velocity field at $N$ distinct spacetime points. Due to the computationally demanding nature of direct numerical simulations of the Navier-Stokes equations (DNS), many questions about high-order velocity moments of multipoint statistics involving many points have been practically inaccessible. Current methods for estimating averaged quantities, RANS and LES, focus mainly on single-point and low-order moments (such as the mean velocity field) and are typically inaccurate for multipoint and high-order velocity moments \citep{pope2000turbulence}. One major challenge of computing multipoint statistics is the closure problem. The governing equation for the $N$-point PDF has terms that depend on the $(N+1)$-point PDF. Although the governing equation for the $(N \rightarrow \infty)$-point probability measure is known to be closed \citep{hopf1952jrma, monin1971vol1and2, kolmogorov1935transformation}, the solutions are beyond the current technical reach.

Recent progress has been made by \cite{SreenivasanYakhot2021moments} where they created a first-principles-based closure for the $n$th-order velocity increment moment equation. This paper obtains the nontrivial pressure closure in the $2$-point velocity increment Hopf equation that is equivalent to the closure of the $n$th-order velocity moment equation from \cite{SreenivasanYakhot2021moments}. This allows the closure to be further generalized to the $N$-point velocity increment Hopf equation containing complete information of the $N$-point probability measure shown in \cref{section:hopf_closure}. Then, in \cref{section:3pt_str_fun_trans}, an example of the closure's capability is demonstrated by determining the functional form of the transition function between the $2$-point structure function and the $3$-point fusion rules for the $3$-point structure function.

\subsection{Background}
\label{sec:SY21_restatement}
This subsection will focus on a short summary of the results and closure procedure done by \cite{SreenivasanYakhot2021moments}. The following notation will be used: various quantities at the point $k$ will be labeled with a superscript, for example the velocity at the $k$th point ($\overline{x}^k$) is $\overline{u}(\overline{x}^k) \equiv \overline{u}^k$. Vector components are represented with a subscript and there is an implied summation over repeated indexes, for example $u^k_j u^k_j = u^k_1 u^k_1+u^k_2 u^k_2+u^k_3 u^k_3$ with no implied summation over $k$. The separation vector is defined as $\overline{r}^k \equiv \overline{x}^k-\overline{x}^0$, where $\overline{x}^0$ is the reference point. The velocity difference is defined as $\delta \overline{u}^k \equiv \overline{u}(\overline{x}^0 + \overline{r}^k) - \overline{u}(\overline{x}^0)$, and the $2$-point characteristic function for the velocity difference $\delta u$ (i.e. the Fourier transform of its PDF) is
\begin{equation}
    \psi(\overline{\omega}^1;\overline{r}^1) \equiv \left\langle e^{i \delta u^1_\mu \omega^1_\mu} \right\rangle.
\end{equation}
For this section take the following notation, due to the presence of only one scale ($r \equiv |\overline{r}|$): $\delta \overline{u}^1 \equiv \delta \overline{u}$, $ \overline{r}^1 \equiv  \overline{r}$, and $\overline{\omega}^1 \equiv \overline{\omega}$. The governing equation for $\psi$ is the $2$-point velocity increment Hopf equation \citep{Yakhot2001strfun, monin1971vol1and2}. For the derivation see appendix \ref{appA_NptHopfeq} and take $N=2$. The equation is
\begin{equation}
    \frac{\partial \psi}{\partial t} - i \frac{\partial^2 \psi}{\partial r_j \partial \omega_j} =
    i I_f + i I_p + i D,
    \label{eq:vel_incr_hopf_eq}
\end{equation}
where
\begin{equation}
    I_f = \left\langle \overline{\omega} \cdot \overline{F} e^{i \overline{\omega} \cdot \delta \overline{u}} \right\rangle,
\end{equation}
\begin{equation}
    I_p = -\left\langle \omega_j \left[ \frac{\partial p(\overline{x}^1)}{\partial x_j^1} - \frac{\partial p(\overline{x}^0)}{\partial x_j^0}   \right] e^{i \overline{\omega} \cdot \delta \overline{u}} \right\rangle,
\end{equation}
\begin{equation}
    D = \nu \left\langle \omega_j \left[ \frac{\partial^2}{\partial x^1_m \partial x_m^1} \delta u_j - \frac{\partial^2}{\partial x^0_m \partial x_m^0} \delta u_j   \right] e^{i \overline{\omega} \cdot \delta \overline{u}} \right\rangle,
\end{equation}
and $\overline{F}$ is the large-scale forcing. In its current form, this equation is unclosed due to the pressure and the viscous terms reliance on information from the statistics of a third point. This hierarchy problem will persist; as seen in appendix \ref{appA_NptHopfeq} the $N$-point characteristic function will require information from involving $(N+1)$-point statistics. Only in the infinite point limit ($N \rightarrow \infty$) is the hierarchy problem circumvented \citep{monin1971vol1and2, hopf1952jrma}, but comes at the cost of heavy computation. Instead, to make progress, \cite{SreenivasanYakhot2021moments} make a closure to the $2$-point case and compute the structure function scaling exponent $\zeta_{2n,0}$. They do this by first making the coordinate transform: $\eta_1 = \sqrt{r_jr_j}$, $\eta_2 = \frac{\omega_j r_j}{r}$, and $\eta_3 = \sqrt{\omega_j \omega_j - (\eta_2)^2}$. The longitudinal and transverse increments are defined below,
respectively:
\begin{equation}
    \delta u \equiv \frac{r_j}{r} \delta u_j,
\end{equation}
\begin{equation}
    \delta v \equiv \frac{\delta u_j}{\eta_3}\left( \omega_j - \eta_2 \frac{r_j}{r} \right).
\end{equation}
The characteristic function becomes
\begin{equation}
    \psi = \left\langle e^{ i \eta_2 \delta u + i \eta_3 \delta v } \right\rangle.
\end{equation}
The $2$-point velocity increment Hopf equation in this coordinate system becomes
\begin{align}
    \frac{\partial \psi}{\partial t} - i \left[ \frac{\partial}{\partial\eta_1} \frac{\partial}{\partial \eta_2} + 
    \frac{d-1}{r} \frac{\partial}{\partial \eta_2} + \frac{\eta_3}{r} \frac{\partial}{\partial \eta_2}  \frac{\partial}{\partial \eta_3} + \frac{(2-d) \eta_2}{r \eta_3} \frac{\partial}{\partial \eta_3} - \frac{\eta_2}{r} \frac{\partial^2}{\partial (\eta_3)^2} \right] \psi =
    \\
    i I_f + i I_p + i D,
\end{align}
where $d$ is the dimensionality of space; in this case $d =3$. Following \cite{SreenivasanYakhot2021moments}, it is assumed the increment scale ($r$) is deep in the inertial range from both limits, i.e. ($\eta_{41} \ll r \ll L$), where $\eta_{41}$ is Kolmogorov dissipation length scale and $L$ is the forcing length scale. This allows us to neglect the direct influences of the viscous and forcing terms ($I_f$ \& $D$). The net effect of the dissipation and viscosity is handled later effectively by the pressure closure. Then assuming the statistics are time stationary, the resulting equation is
\begin{equation}
         -\left[ \frac{\partial}{\partial\eta_1} \frac{\partial}{\partial \eta_2} + 
    \frac{d-1}{r} \frac{\partial}{\partial \eta_2} + \frac{\eta_3}{r} \frac{\partial}{\partial \eta_2}  \frac{\partial}{\partial \eta_3} + \frac{(2-d) \eta_2}{r \eta_3} \frac{\partial}{\partial \eta_3} - \frac{\eta_2}{r} \frac{\partial^2}{\partial (\eta_3)^2} \right] \psi = I_p.
    \label{eq:unclosed_2pt_hopf_simple}
\end{equation}
The $2$-point structure function is defined as $S_{n,m} \equiv \left\langle \delta u^n \delta v^m \right\rangle$. To get the equation for the longitudinal structure function ($S_{2n,0}$), the following operator is defined
\begin{equation}
    O \equiv \frac{\partial}{\eta_3} \left( \frac{\partial}{\partial \eta_2} \right)^{2n-1} \eta_3.
\end{equation}
The operator $O$ is applied to \cref{eq:unclosed_2pt_hopf_simple} and then the following limits are taken: $\eta_2=\eta_3\rightarrow0$ and $\eta_1=r$. The resulting equation is \citep{SreenivasanYakhot2021moments, Yakhot2001strfun, hill2001strfun}
\begin{align}
    \label{eq:unclosed_str_fun_eq}
    \frac{\partial S_{2n,0}}{\partial r } + \frac{d-1}{r} S_{2n,0} - \frac{(d-1)(2n-1)}{r}S_{2n-2,2} =
    \\
    -(2n-1)\left\langle \delta_r\frac{\partial p}{\partial x} \left( \delta u \right)^{2n-1} \right\rangle,
\end{align}
which is unclosed. \cite{SreenivasanYakhot2021moments} make a series of physical arguments to justify the the following closure to the pressure term
\begin{align}
    \frac{\partial S_{2n,0}}{\partial r } + \frac{d-1}{r} S_{2n,0} - \frac{(d-1)(2n-1)}{r}S_{2n-2,2} \approx
    \\
    -(2n-1) \left\langle \left[ -\frac{\partial \delta u}{\partial t} - \frac{1}{2}\frac{\partial (\delta u)^2}{\partial r} + \frac{\partial}{\partial r} \nu_T \frac{\partial \delta u}{\partial r} + \mathcal{O} (\delta v)^2\right] (\delta u)^{2n-2}\right\rangle
\end{align}
It is said that $\frac{\partial \delta u}{\partial t} \sim \frac{\partial (\delta u)^2}{\partial r}$ and the constants $a \approx \left\langle  -\frac{\partial \delta u}{\partial t} -\frac{1}{2} \frac{\partial (\delta u)^2}{\partial r}\right \rangle$ and $b \approx \left\langle (\delta v)^2 (\delta u)^{2n-2} \right\rangle$. The equation becomes
\begin{align}
    \label{eq:closed_str_fun_eq}
    \frac{\partial S_{2n,0}}{\partial r } + \frac{d-1}{r} S_{2n,0} - \frac{(d-1)(2n-1)}{r}S_{2n-2,2} =
    \\
    (2n-1)\frac{a}{n} \frac{\partial S_{2n,0}}{\partial r} - \frac{2n-1}{r} b S_{2n-2,2}.
\end{align}
To find the scaling exponents the ansatz $S_{n,m} = A_{n,m}r^{\zeta_{n,m}}$ is made. Then it is assumed that $\zeta_{2n,0} \approx \zeta_{2n-2,2}$. Plugging this in and solving for $\zeta_{2n,0}$ gives
\begin{equation}
    \zeta_{2n,0} = \frac{(2n-1)(2-b)\frac{A_{2n-2,2}}{A_{2n,0}}-2}{(1-2a)n+a}n.
\end{equation}
The coefficients $a$ and $b$ are determined to satisfy $\zeta_{3,0}=1$ and from the effective eddy viscosity $\nu_T$. \cite{SreenivasanYakhot2021moments} determine the constants to be $a=0.475$ and $b = 0.225$. Furthermore, it is assumed $\frac{A_{2n-2,2}}{A_{2n,0}} \approx \frac{4}{3}\frac{1}{2n-1}$. The final expression found by \cite{SreenivasanYakhot2021moments} is
\begin{equation}
    \zeta_{2n,0} = \frac{0.366n}{0.05n + 0.475}.
\end{equation}

\section{Structure Function Hopf Equation Closure and Generalization}
\label{section:hopf_closure}

Now this section will provide a closure to the velocity increment Hopf equation (\cref{eq:vel_incr_hopf_eq}) and then prove an equivalence between this closure and the closure to the structure function equation provided by \cite{SreenivasanYakhot2021moments} that was recapped in \cref{sec:SY21_restatement}. As stated earlier in \cref{sec:SY21_restatement}, the $2$-point velocity difference Hopf equation in transformed coordinates is
\begin{align}
    \label{eq:unclosed_2pt_hopf_transformed}
    \frac{\partial \psi}{\partial t} - i \left[ \frac{\partial}{\partial\eta_1} \frac{\partial}{\partial \eta_2} + 
    \frac{d-1}{r} \frac{\partial}{\partial \eta_2} + \frac{\eta_3}{r} \frac{\partial}{\partial \eta_2}  \frac{\partial}{\partial \eta_3} + \frac{(2-d) \eta_2}{r \eta_3} \frac{\partial}{\partial \eta_3} - \frac{\eta_2}{r} \frac{\partial^2}{\partial (\eta_3)^2} \right] \psi =
    \\
    i I_f + i I_p + i D.
\end{align}
The characteristic function can be written as
\begin{equation}
    \psi = \left\langle e^{i \overline{\omega} \cdot \delta \overline{u}} \right\rangle
\end{equation}
In this particular coordinate system the characteristic function can be equivalently expressed as, 
\begin{equation}
    \psi = \left\langle e^{ i \eta_2 \delta_ru + i \eta_3 \delta_r v } \right\rangle.
\end{equation}
Following \cite{SreenivasanYakhot2021moments} it is assumed the increment scale $r$ is deep in the inertial range from both limits, i.e. ($\eta_{41} \ll r \ll L$), where $\eta_{41}$ is Kolmogorov dissipation length scale and $L$ is the forcing length scale. Therefore, the direct influences of the viscous and forcing terms ($I_f$ \& $D$) are neglected. It is assumed  the statistics do not depend on time leaving,
\begin{equation}
    \label{eq:closed_2pt_hopf}
         -\left[ \frac{\partial}{\partial\eta_1} \frac{\partial}{\partial \eta_2} + 
    \frac{d-1}{r} \frac{\partial}{\partial \eta_2} + \frac{\eta_3}{r} \frac{\partial}{\partial \eta_2}  \frac{\partial}{\partial \eta_3} + \frac{(2-d) \eta_2}{r \eta_3} \frac{\partial}{\partial \eta_3} - \frac{\eta_2}{r} \frac{\partial^2}{\partial (\eta_3)^2} \right] \psi = I_p,
\end{equation}
which is unclosed in its current form due to the pressure term. The following closure is proposed for the pressure
\begin{equation}
    \label{eq:2pt_hopf_pressure_closure}
    I_p = -a \frac{\partial}{\partial \eta_1} \frac{\partial}{\partial \eta_2} M_{s \rightarrow \eta_2}^{-1} \left[ \frac{2(s+1)}{s} M_{  \eta_2 \rightarrow s} (\psi) \right] + \left( \frac{b}{r} \frac{\partial^2}{\partial (\eta_3)^2} \eta_2 \right)\psi.
\end{equation}
The constants $a$ and $b$ are the same as in \cref{sec:SY21_restatement}. The closure makes use of the Mellin and inverse Mellin transforms defined by
\begin{equation}
    \Tilde{\phi}(s) = M_{  \eta_2 \rightarrow s} (\phi) \equiv \int_0^{\infty} d \eta_2 \eta_2^{s-1} \phi(\eta_2),
\end{equation}
\begin{equation}
    \phi(\eta_2) = M_{s \rightarrow \eta_2}^{-1}(\Tilde{\phi}) \equiv \frac{1}{2 \pi i} \int_{\gamma-i\infty}^{\gamma+i\infty} ds \eta_2^{-s} \Tilde{\phi}(s).
\end{equation}
To demonstrate an equivalence between this closure and the closure from \cite{SreenivasanYakhot2021moments}, the governing equation for the structure functions are derived by first applying the operator $\frac{\partial}{\partial \eta_3} \left( \frac{\partial}{\partial \eta_2} \right)^{2n-1} \eta_3$ to \cref{eq:closed_2pt_hopf} and then taking $\eta_2=\eta_3\rightarrow0$ and $\eta_1=r$. The mathematical procedure is laid out in appendix \ref{appB_str_from_hopf_closure}, but the result is:
\begin{align}
    \frac{\partial S_{2n,0}}{\partial r } + \frac{d-1}{r} S_{2n,0} - \frac{(d-1)(2n-1)}{r}S_{2n-2,2} =
    \\
    (2n-1)\frac{a}{n} \frac{\partial S_{2n,0}}{\partial r} - \frac{2n-1}{r} b S_{2n-2,2}.
\end{align}
This is exactly the closed structure function equation (\cref{eq:closed_str_fun_eq}) from \cref{sec:SY21_restatement} that contains the same scaling exponents ($\zeta_{2n,0}$) as \cite{SreenivasanYakhot2021moments}. This demonstrates the equivalence between the closure to the velocity increment Hopf equation provided here and to the structure function equation closure found by \cite{SreenivasanYakhot2021moments}.

\subsection{$N$-point SY Hopf Closure}
There exists a natural extension of the $2$-point pressure closure, provided in the previous section, to a pressure closure for the $N$-point velocity increment Hopf equation. There are an infinite number of possible closures and extensions of the $2$-point closure for the $N$-point closure. The particular functional form of the $N$-point closure in this paper is chosen for its simplicity (without being too simple) and its natural mathematical flavor that appears to unify a picture between the $2$-point and $N$-point closures. Additionally, any dimensional reduction back to the $2$-point case returns to the $2$-point Hopf closure in \cref{section:hopf_closure} as expected.
To start, the unclosed $N$-point velocity without the viscous and forcing terms is repeated here from appendix \ref{appA_NptHopfeq}, where it is derived:
\begin{equation}
    \label{eq:unclosed_multipt_hopf}
    \frac{\partial \psi}{\partial t} - i \sum_{k=1}^{N-1} \frac{\partial^2 \psi}{\partial r_j^k \partial \omega_j^k} =  i \sum_{k=1}^{N-1} I_p^k,
\end{equation}
where
\begin{equation}
    I_p^k = - \omega_j^k \left\langle \left( \frac{\partial p(\overline{x}^k)}{\partial x_j^k} - \frac{\partial p(\overline{x}^0)}{\partial x_j^0} \right) \text{exp}\left( i \sum_{\alpha = 1}^{N-1} \omega_{\mu}^{\alpha} \delta u_{\mu}^{\alpha} \right) \right\rangle.
\end{equation}
The multipoint characteristic function is
\begin{equation}
\psi(\overline{\omega}^1,...,\overline{\omega}^{N-1};\overline{r}^1,...,\overline{r}^{N-1}) = \left\langle \text{exp}\left( i \sum_{\alpha = 1}^{N-1} \omega_{\mu}^{\alpha} \delta u_{\mu}^{\alpha} \right) \right\rangle.
\end{equation}
The same transformation as earlier is used but for the multipoint case : $\eta_1^k = \sqrt{r_j^kr_j^k}$, $\eta_2^k = \frac{\omega_j^k r_j^k}{r^k}$, and $\eta_3^k = \sqrt{\omega_j^k \omega_j^k - (\eta_2^k)^2}$. The $k$th increment scale is defined as $r^k \equiv |\overline{r}^k|$. The multipoint longitudinal and transverse increments are defined below,
respectively:
\begin{equation}
    \delta u^k \equiv \frac{r_j^k}{r^k} \delta u_j^k,
\end{equation}
\begin{equation}
    \delta v^k \equiv \frac{\delta u_j^k}{\eta_3^k}\left( \omega_j^k - \eta_2^k \frac{r_j^k}{r^k} \right).
\end{equation}
The characteristic function is reformulated, in terms of the transformed coordinates, as
\begin{equation}
    \psi = \left\langle \text{exp}\left( i \sum_{\alpha = 1}^{N-1} \left( \eta_2^{\alpha} \delta u^{\alpha} + \eta_3^{\alpha} \delta v^{\alpha} \right) \right) \right\rangle.
\end{equation}
Since each term in the governing equation (\cref{eq:unclosed_multipt_hopf}) only depends on one scale, each transformation is independent and can be performed by applying the $2$-point case transformation independently for each $k$ up to $N-1$. The proposed closure, in the transformed coordinates, for the pressure term is:
\begin{equation}
    I_p^k =  -a \frac{\partial}{\partial \eta^k_1} \frac{\partial}{\partial \eta^k_2} M_{s \rightarrow \eta^k_2}^{-1} \left[ \frac{2(s+1)}{s} M_{  \eta^k_2 \rightarrow s} (\psi) \right] + \left( \frac{b}{r} \frac{\partial^2}{\partial (\eta_3^k)^2} \eta_2^k \right)\psi.
\end{equation}
In total, the $N$-point velocity increment Hopf equation in the transformed coordinates becomes:
\begin{align}
\frac{\partial \psi}{\partial t} - i \sum_{k=1}^{N-1} \left[ \frac{\partial}{\partial\eta_1^k} \frac{\partial}{\partial \eta_2^k} + 
    \frac{d-1}{r^k} \frac{\partial}{\partial \eta^k_2} + \frac{\eta^k_3}{r^k} \frac{\partial}{\partial \eta^k_2}  \frac{\partial}{\partial \eta^k_3} + \frac{(2-d) \eta^k_2}{r^k \eta^k_3} \frac{\partial}{\partial \eta^k_3} - \frac{\eta^k_2}{r^k} \frac{\partial^2}{\partial (\eta_3^k)^2} \right] \psi =
\\
    -i \sum_{k=1}^{N-1} \left\{ 
    a \frac{\partial}{\partial \eta^k_1} \frac{\partial}{\partial \eta^k_2} M_{s \rightarrow \eta^k_2}^{-1} \left[ \frac{2(s+1)}{s} M_{  \eta^k_2 \rightarrow s} (\psi) \right] - \left( \frac{b}{r} \frac{\partial^2}{\partial (\eta_3^k)^2} \eta_2^k \right)\psi \right\}.
\label{eq:closed_multipt_hopf_transformed}
\end{align}
The equation above is closed. The equation remains closed whether or not the forcing terms are included, but is unclosed if the viscous terms are reintroduced. However, both the forcing and viscous terms are neglected for inertial range calculations \citep{SreenivasanYakhot2021moments}. For completeness, an inverse transformation can be applied to \cref{eq:closed_multipt_hopf_transformed} to have the equations in the standard coordinates. The equation is as follows:
\begin{align}
\frac{\partial \psi}{\partial t} - i \sum_{k=1}^{N-1} \frac{\partial^2 \psi}{\partial r^k_j \partial \omega^k_j} =
\\
    -i \sum_{k=1}^{N-1} \Biggl\{
    a \frac{\partial}{\partial |\overline{r}^k|} \frac{\partial}{\partial \omega^k_m \frac{r^k_m}{|\overline{r}^k|}} M_{s \rightarrow \omega^k_m \frac{r^k_m}{|\overline{r}^k|}}^{-1} \left[ \frac{2(s+1)}{s} M_{  \omega^k_m \frac{r^k_m}{|\overline{r}^k|} \rightarrow s} (\psi) \right]
    \\
    - \left( \frac{b}{r} \left(\frac{\partial}{\partial \sqrt{\omega^k_m \omega^k_m - \left(\omega^k_m \frac{r^k_m}{|\overline{r}^k|} \right)^2}}\right)^2 \omega^k_{\mu} \frac{r^k_{\mu}}{|\overline{r}^k|} \right)\psi \Biggl\}.
\label{eq:closed_multipt_hopf}
\end{align}

\section{3-point Structure Function Transition}
\label{section:3pt_str_fun_trans}

\begin{figure}
    \centering
    \includegraphics[width=0.95\linewidth]{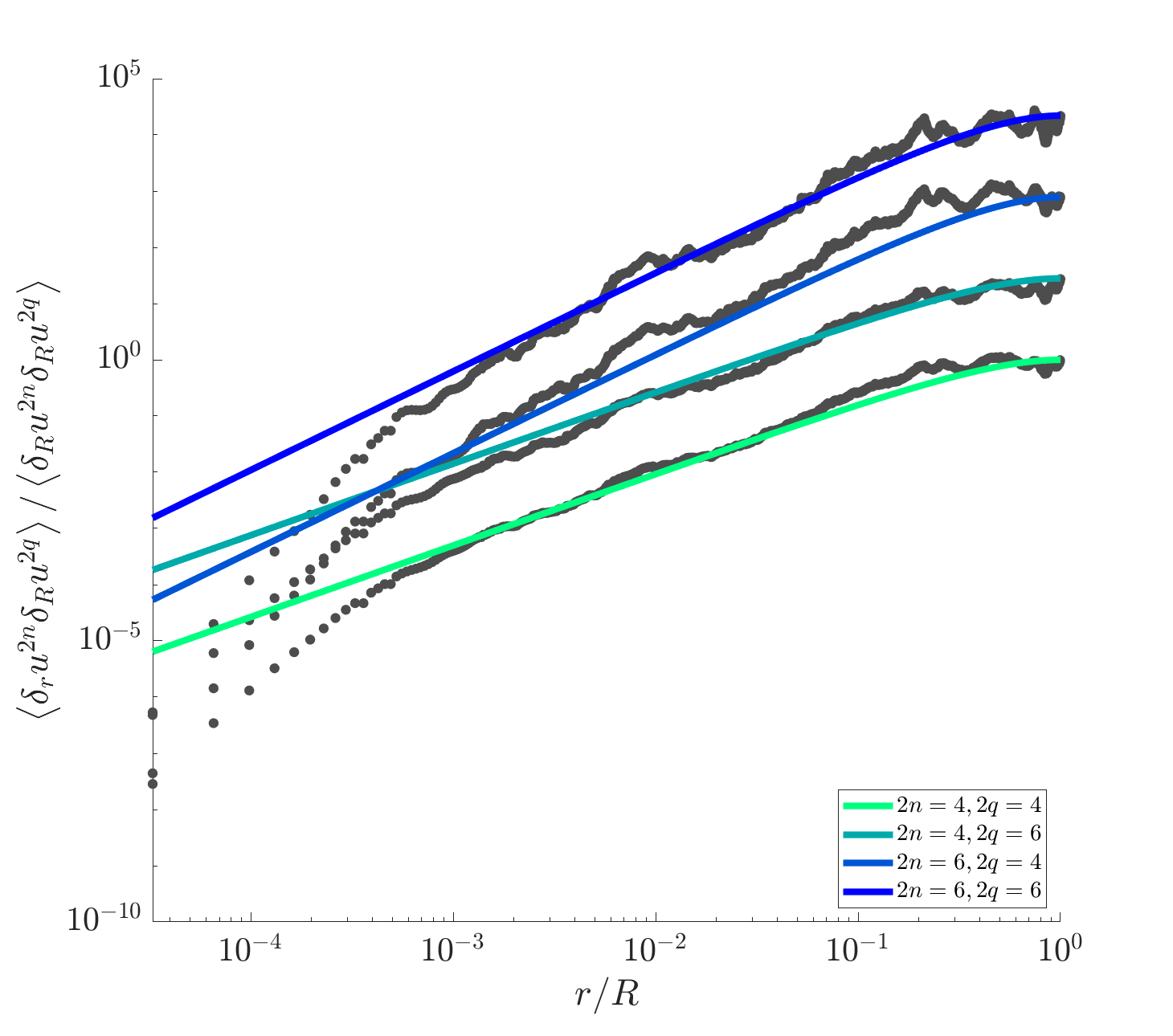}
    \caption{The normalized even $3$-point structure function plotted for various $n$ and $p$ using the analytical derivation from the Hopf closure (solid lines) and from about $80,000$ samples of the \texttt{iso32768} DNS dataset from the Johns Hopkins Turbulence Database (JHTDB), for details about the simulation see \citet{JHUTurbulence, Yeung2025ExascaleTurbulence}.}
    \label{fig:3pt_str_fun}
\end{figure}

To demonstrate and test the capability of the multipoint closure, this section will derive a transition function for the $3$-point structure function between the known limits. In this case, the characteristic function is
\begin{equation}
    \psi = \left\langle \text{exp}\left( i \sum_{\alpha = 1}^{2} \left( \eta_2^{\alpha} \delta u^{\alpha} + \eta_3^{\alpha} \delta v^{\alpha} \right) \right) \right\rangle.
\end{equation}
The governing equation for $\psi$ is the closed multipoint velocity increment Hopf equation (\cref{eq:closed_multipt_hopf_transformed}) in transformed coordinates for $N=3$:
\begin{align}
\frac{\partial \psi}{\partial t} - i \sum_{k=1}^2 \left[ \frac{\partial}{\partial\eta_1^k} \frac{\partial}{\partial \eta_2^k} + 
    \frac{d-1}{r^k} \frac{\partial}{\partial \eta^k_2} + \frac{\eta^k_3}{r^k} \frac{\partial}{\partial \eta^k_2}  \frac{\partial}{\partial \eta^k_3} + \frac{(2-d) \eta^k_2}{r^k \eta^k_3} \frac{\partial}{\partial \eta^k_3} - \frac{\eta^k_2}{r^k} \frac{\partial^2}{\partial (\eta_3^k)^2} \right] \psi =
\\
    -i \sum_{k=1}^{2} \left\{ 
    a \frac{\partial}{\partial \eta^k_1} \frac{\partial}{\partial \eta^k_2} M_{s \rightarrow \eta^k_2}^{-1} \left[ \frac{2(s+1)}{s} M_{  \eta^k_2 \rightarrow s} (\psi) \right] - \left( \frac{b}{r} \frac{\partial^2}{\partial (\eta_3^k)^2} \eta_2^k \right)\psi \right\}.
\end{align}
Once again, following \cite{SreenivasanYakhot2021moments}, it is assumed that all increment lengths are well into the inertial range, allowing the viscous and forcing terms to be neglected. Assuming the statistics are not a function of time gives
\begin{align}
  \sum_{k=1}^2 \left[ \frac{\partial}{\partial\eta_1^k} \frac{\partial}{\partial \eta_2^k} + 
    \frac{d-1}{r^k} \frac{\partial}{\partial \eta^k_2} + \frac{\eta^k_3}{r^k} \frac{\partial}{\partial \eta^k_2}  \frac{\partial}{\partial \eta^k_3} + \frac{(2-d) \eta^k_2}{r^k \eta^k_3} \frac{\partial}{\partial \eta^k_3} - \frac{\eta^k_2}{r^k} \frac{\partial^2}{\partial (\eta_3^k)^2} \right] \psi =
\\
    \sum_{k=1}^{2} \left\{ 
    a \frac{\partial}{\partial \eta^k_1} \frac{\partial}{\partial \eta^k_2} M_{s \rightarrow \eta^k_2}^{-1} \left[ \frac{2(s+1)}{s} M_{  \eta^k_2 \rightarrow s} (\psi) \right] - \left( \frac{b}{r} \frac{\partial^2}{\partial (\eta_3^k)^2} \eta_2^k \right)\psi \right\}.
    \label{eq:reduced_3pt_hopf}
\end{align}
For brevity, define the two increment scales as $r\equiv r^1$ and $R \equiv r^2$. The $3$-point structure function is defined as
\begin{equation}
    S_{n,m,q,s} \equiv \left\langle (\delta_r u)^n (\delta_r v)^m (\delta_R u)^p (\delta_R v)^s \right\rangle,
\end{equation}
where
\begin{equation}
    \delta_r u \equiv \frac{r_j^1}{r}\delta u^1_j,
\end{equation}
\begin{equation}
    \delta_R u \equiv \frac{r_j^2}{R}\delta u^2_j,
\end{equation}
\begin{equation}
    \delta_r v \equiv \frac{\delta u_j^1}{\eta_3^1} \left( \omega^1_j - \eta_2^1 \frac{r_j^1}{r} \right),
\end{equation}
\begin{equation}
    \delta_R v \equiv \frac{\delta u_j^2}{\eta_3^2} \left( \omega^2_j - \eta_2^2 \frac{r_j^2}{R} \right).
\end{equation}
The $3$-point structure function can be obtained from the characteristic function evaluating the following expression
\begin{equation}
    S_{n,m,q,s} = \left. \left( \frac{1}{i}\frac{\partial}{\partial \eta_2^1} \right)^n  \left( \frac{1}{i}\frac{\partial}{\partial \eta_3^1} \right)^m  \left( \frac{1}{i}\frac{\partial}{\partial \eta_2^2} \right)^q  \left( \frac{1}{i}\frac{\partial}{\partial \eta_3^2} \right)^s \psi \right|_{\eta_2^1=\eta_3^1=\eta_2^2=\eta_3^2=0,\eta_1^1 = r,\eta^2_1=R}.
\end{equation}
To obtain the governing equation for the $3$-point structure function the operator $O$ (defined below) is applied to \cref{eq:reduced_3pt_hopf} and evaluated at $\eta_2^1=\eta_3^1=\eta_2^2=\eta_3^2=0$, $\eta_1^1 = r$, and $\eta^2_1=R$
\begin{equation}
    O \equiv \frac{\partial}{\partial \eta_3^1} \left( \frac{\partial}{\partial \eta_2^1} \right)^{2n-1} \eta_3^1 \frac{\partial}{\partial \eta_3^2} \left( \frac{\partial}{\partial \eta_2^2} \right)^{2p-1} \eta_3^2.
\end{equation}
The form of $O$ comes from a natural extension of the operator used by \cite{SreenivasanYakhot2021moments} in \cref{sec:SY21_restatement}. The resulting equation is
\begin{align}
    \frac{\partial}{\partial r} S_{2n,0,2q-1,0} + \frac{d-1}{r} S_{2n,0,2q-1,0} - \frac{(d-1)(2n-1)}{r}S_{2n-2,2,2q-1,0} + 
    \\
    \frac{\partial}{\partial R} S_{2n-1,0,2q,0} + \frac{d-1}{R} S_{2n-1,0,2q,0} - \frac{(d-1)(2q-1)}{R}S_{2n-1,0,2q-2,2} =
    \\
    (2n-1)\frac{a}{n}\frac{\partial}{\partial r}S_{2n,0,2q-1,0} - \frac{2n-1}{r}b S_{2n-2,2,2q-1,0}
    \\
    +(2q-1)\frac{a}{q}\frac{\partial}{\partial R}S_{2n-1,0,2q,0} - \frac{2q-1}{R}b S_{2n-1,0,2q-2,2}.
    \label{eq:3pt_structure_function}
\end{align}
In a similar procedure to \cref{sec:SY21_restatement}, the following functional form for the structure function is assumed
\begin{equation}
    \left\langle (\delta_r u)^n (\delta_r v)^m (\delta_R u)^q (\delta_R v)^s\right\rangle = A_{n,m,q,s} R^{\zeta_{n+q,m+s}} F_{n,m,q,s}\left(\frac{r}{R}\right),
    \label{eq:3pt_structure_function_assumed_form}
\end{equation}
where $A_{n,m,q,s}$ is constant with $\frac{r}{R}$, $\zeta_{n,m}$ is the usual $2$-point structure function scaling exponent found in \cref{sec:SY21_restatement}, and $F_{n,m,q,s}\left(\frac{r}{R}\right)$ is a transition function that must be solved for. Now the following limits exist for $S_{n,m,q,s}$ , which imposes limits for $F_{n,m,q,s}$: in the limit $r \approx R$, the two scales should combine to become the regular $2$-point structure function. That is 
\begin{equation}
    \left\langle (\delta_R u)^n (\delta_R v)^m (\delta_R u)^q (\delta_R v)^s\right\rangle = \left\langle (\delta_R u)^{n+q} (\delta_R v)^{m+s} \right\rangle,
\end{equation}
therefore,
\begin{equation}
    F_{n,m,q,s}(1) = \frac{A_{n+q,m+s,0,0}}{A_{n,m,q,s}}.
    \label{eq:F_bc1}
\end{equation}
Additionally, it is expected that, in the limit $r \rightarrow R$, the $3$-point structure will have constant scaling in $r$: in other words $F \sim r^0$. This leads to the boundary condition
\begin{equation}
    \left. \frac{d \text{log}(F_{n,m,q,s})}{d \text{log} \left( \frac{r}{R} \right)} \right|_{\frac{r}{R} = 1} = 0.
    \label{eq:F_bc2}
\end{equation}
In the other extreme if $r<<R$ then the fusion rules are expected for $n > 1$ and $q > 1$ \citep{Benzi1998fusion, Benzi1999fusion, friedrich2018fusion}. For the longitudinal case ($m=0$, $s=0$) that is
\begin{equation}
    \left\langle (\delta_R u)^n (\delta_R v)^0 (\delta_R u)^q (\delta_R v)^0\right\rangle = B_{n,q}\frac{\langle  (\delta_r u)^n \rangle}{\langle (\delta_R u)^n \rangle} \langle (\delta_R u)^{n+q} \rangle.
\end{equation}
As a result, in the limit $\frac{r}{R} \rightarrow 0$
\begin{equation}
    F_{n,0,q,0}\left(\frac{r}{R}\right) = B_{n,q} \frac{A_{n+q,0,0,0}}{A_{n,0,q,0}} \left( \frac{r}{R} \right)^{\zeta_{n,0}}.
    \label{eq:F_bc3}
\end{equation}
Plug in the ansatz (\cref{eq:3pt_structure_function_assumed_form}) into the governing equation for the $3$-point structure function (\cref{eq:3pt_structure_function}) and make the following assumptions:
\begin{equation}
    \zeta_{2n,2} \approx \zeta_{2n+2,0},
    \label{eq:zeta_assumption}
\end{equation}
\begin{equation}
    F_{2n-2,2,2q,0} \approx F_{2n,0,2q,0},
    \label{eq:F_assumption_1}
\end{equation}
\begin{equation}
    F_{2n-1,0,2q,0} \approx F_{2n,0,2q-1,0}.
    \label{eq:F_assumption_2}
\end{equation}
The first assumption \cref{eq:zeta_assumption} was used in \cite{SreenivasanYakhot2021moments} for the calculation of the scaling exponents $\zeta_{2n,0}$. Assumptions \cref{eq:F_assumption_1} and \cref{eq:F_assumption_2} are similar in spirit to \cref{eq:F_assumption_1}. Other approximations can be made which will lead to different final results. It is important to note that these assumptions are made only to extract information from \cref{eq:reduced_3pt_hopf}, which in principle can be solved (or approximated numerically) without making these assumptions since the equation is closed. The result is the following ODE for $F_{2n,0,2q-1,0}$:
\begin{equation}
    \left( c_1 \left( \frac{r}{R} \right) + c_2 \left( \frac{r}{R} \right)^2\right) F_{2n,0,2q-1,0}' + \left( c_3 \left( \frac{r}{R} \right) + c_4 \right) F_{2n,0,2q-1,0} = 0.
\end{equation}
The solution to the ODE is 
\begin{equation}
    \label{eq:F_functional_form}
    F_{2n,0,2q-1,0}\left( \frac{r}{R} \right) = K \left( \frac{r}{R} \right)^{-c_4/c_1} \left(  c_1 + c_2  \frac{r}{R} \right)^{c_4/c_1 - c_3/c_2}.
\end{equation}
Applying the boundary conditions (\cref{eq:F_bc1}, \cref{eq:F_bc2}, and \cref{eq:F_bc3}) give expressions for the constants $c_1$, $c_2$, $c_3$, $c_4$, and $K$:
\begin{equation}
    c_1 = \left(2-(2q-1)\frac{a}{q} \right),
\end{equation}
\begin{equation}
    c_2 = \left((2q-1)\frac{a}{q} -1 \right),
\end{equation}
\begin{align}
    c_3 = \zeta_{2n,0} \left(2-(2q-1)\frac{a}{q} \right),
\end{align}
\begin{equation}
    c_4 = -\zeta_{2n,0} \left(2-(2q-1)\frac{a}{q} \right),
\end{equation}
\begin{equation}
    K = \frac{A_{2n+2q-1,0,0,0}}{A_{2n,0,2q-1,0}}.
\end{equation}
Additionally, the theory makes the following prediction for the fusion rule coefficient:
\begin{equation}
    B_{2n,2q-1} = c_1^{c_4/c_1 - c_3/c_2}.
\end{equation}
It is important to note that the current calculation does not predict the value of $K$. However, since characteristic function equation is closed (\cref{eq:reduced_3pt_hopf}) in principle $K$ can be calculated with the current theory, but there is not, at the present moment, a known method to analytically determine $K$. This is left to future works. Interesting enough is that \cref{eq:F_functional_form} appears to be a type of Batchelor interpolation: a technique commonly used in turbulence modeling as a transition between two regimes \citep{pope2000turbulence}. This is interesting because this was not designed in but rather came as a consequence of earlier derivations. To give an early test to the validity of this theory, the prediction of the normalized $3$-point structure function, $\frac{\left\langle \delta_ru^{2n} \delta_Ru^{2p} \right\rangle}{\left\langle \delta_Ru^{2n} \delta_Ru^{2p} \right\rangle}$ made from \cref{eq:3pt_structure_function_assumed_form} and \cref{eq:F_functional_form} is plotted (lines) alongside DNS data in \cref{fig:3pt_str_fun}. The DNS data comes from \texttt{iso32768} dataset from the Johns Hopkins Turbulence Database (JHTDB), for details about the simulation see \citet{JHUTurbulence, Yeung2025ExascaleTurbulence}. Even with the noisy data from a low number of samples, the data seem to support the results of the theory for separations in the inertial range. Future in-depth analysis must be done comparing the prediction made here to DNS data. 

\section{Conclusion}

The central contribution of this work is that it extends the first-principles-based closure by \cite{SreenivasanYakhot2021moments} to the equation governing $N$-point statistics. First, a recap of the calculation of the $2$-point structure function scaling exponent by \cite{SreenivasanYakhot2021moments} is provided. Their closure of the pressure term of the structure function governing equation is then extended to an equivalent closure of the pressure term in the $2$-point velocity increment Hopf equation. A natural extension of this closure is used to close the $N$-point velocity increment Hopf equation. 
To demonstrate the capabilities of this method, the $3$-point velocity increment transition function between the known $2$-point structure function and the $3$-point fusion rules is analytically derived from the closed $(N=3)$-point velocity increment Hopf equation. The analytical solution takes the form of a Batchelor interpolation and is plotted against DNS data. The data seem to support the validity of the analytical function for separations in the inertial range.
The results of this work are tools to hopefully analytically probe deeper into a more complex set of statistics characterizing turbulence. It is important to note that this theory has been developed for statistically stationary, homogeneous, isotropic, and incompressible turbulence for statistical quantities within the inertial range. At the moment, it is not immediately clear how these results can be generalized for more complex turbulent flows. Ideally, future works can continue to use this first-principles-based method to derive various accurate multipoint statistical quantities to deepen our understanding of turbulence. 

\section*{Acknowledgments}
I would like to thank my PhD advisor Perry L. Johnson for guiding me; without his guidance throughout my PhD the present work would not have been possible. Additionally, I would like to thank Michael Wilczek for his role as a mentor, as well as his group at University of Bayreuth: Lukas Bentkamp and Gabriel B. Apolin\'ario for their helpful discussions. Finally, I would like to thank Katepalli R. Sreenivasan and Victor Yakhot for laying the groundwork that made this possible.

\begin{bmhead}[Author ORCID:]
https://orcid.org/0009-0006-4171-4415
\end{bmhead}
\begin{bmhead}[]
14FF5BD71B7C049B30AD1D85BDF329A6DCFDE9624F68752EEBE196CDE344E8BA
\end{bmhead}

\begin{appen}

\section{$N$-point velocity difference Hopf equation}\label{appA_NptHopfeq}

This appendix will derive the $N$-point velocity difference equation. This derivation is general for any choice of the natural number $N$; it provides the governing equation for the $N$-point velocity difference characteristic function. \cite{hill2001strfun} provides the velocity difference momentum governing equation is
\begin{equation}
        \frac{\partial \delta u^k_i}{\partial t} + U^k_j \frac{\partial \delta u^k_i}{\partial y_j^k} + \delta u_j^k \frac{\partial \delta u^k_i}{\partial r^k_j } =
    -\left[ \frac{\partial p(\overline{x}^k)}{\partial x^k_i} -  \frac{\partial p(\overline{x}^0)}{\partial x^0_i}\right] + \nu \left( \frac{\partial^2}{\partial x_j^k \partial x_j^k} \delta u^k_i - \frac{\partial^2}{\partial x_j^0 \partial x_j^0} \delta u^0_i  \right),
\end{equation}
where $y^k_i \equiv \frac{1}{2}(x^k_i + x^0_i)$ and $U_i^k \equiv \frac{1}{2}(u_i^k + u_i^0 )$. Conversation of mass gives
\begin{equation}
    \label{eq:mass_conv_1}
    \frac{\partial \delta u_j^k}{\partial r_j^k} = 0,
\end{equation}
\begin{equation}
    \label{eq:mass_conv_2}
    \frac{\partial U_j^k}{\partial y_j^k} = 0.
\end{equation}
The $N$-point velocity increment characteristic function is
\begin{equation}
    \psi(\overline{\omega}^1,...,\overline{\omega}^{N-1};\overline{r}^1,...,\overline{r}^{N-1}) \equiv \left\langle \text{exp}\left\{ i \sum_{\alpha =1}^{N-1} \omega^{\alpha}_\mu \delta u^\alpha_\mu \right\} \right\rangle.
\end{equation}
Then
\begin{equation}
    \frac{\partial \psi}{\partial t} = \left\langle \frac{\partial}{\partial t} \text{exp}\left\{ i \sum_{\alpha =1}^{N-1} \omega^{\alpha}_\mu \delta u^\alpha_\mu \right\} \right\rangle
\end{equation}
\begin{equation}
     = \left\langle i \sum_{k=1}^{N-1} \omega_i^k \frac{\partial \delta u_i^k}{\partial t} \text{exp}\left\{ i \sum_{\alpha =1}^{N-1} \omega^{\alpha}_\mu \delta u^\alpha_\mu \right\} \right\rangle
\end{equation}
\begin{align}
    = \Biggl \langle i \sum_{k=1}^{N-1} \omega_i^k \Biggl\{ - U^k_j \frac{\partial \delta u^k_i}{\partial y_j^k} - \delta u_j^k \frac{\partial \delta u^k_i}{\partial r^k_j } 
    \\
    -\left[ \frac{\partial p(\overline{x}^k)}{\partial x^k_i} -  \frac{\partial p(\overline{x}^0)}{\partial x^0_i}\right] + \nu \left( \frac{\partial^2}{\partial x_j^k \partial x_j^k} \delta u^k_i - \frac{\partial^2}{\partial x_j^0 \partial x_j^0} \delta u^0_i  \right)  \Biggl\} \text{exp}\left\{ i \sum_{\alpha =1}^{N-1} \omega^{\alpha}_\mu \delta u^\alpha_\mu \right\} \Biggr \rangle.
\end{align}
Focus in on the following particular term
\begin{equation}
     \left\langle i \sum_{k=1}^{N-1} \omega_i^k \left( - U^k_j \frac{\partial \delta u^k_i}{\partial y_j^k} \right) \text{exp}\left\{ i \sum_{\alpha =1}^{N-1} \omega^{\alpha}_\mu \delta u^\alpha_\mu \right\} \right\rangle.
\end{equation}
Using mass conservation (\cref{eq:mass_conv_1} and \cref{eq:mass_conv_2}) this can be rewritten as
\begin{equation}
     -\sum_{k=1}^{N-1} \frac{\partial }{\partial y_j^k} \left\langle     U^k_j   \text{exp}\left\{ i \sum_{\alpha =1}^{N-1} \omega^{\alpha}_\mu \delta u^\alpha_\mu \right\} \right\rangle.
     \label{eq:hopf_zero_term}
\end{equation}
In homogeneous isentropic turbulence, it is assumed $\frac{\partial}{\partial y_j^k} \langle ...\rangle = 0$. Therefore, the term given by \cref{eq:hopf_zero_term} in the Hopf equation can be neglected. Next the convective term can be written as
\begin{equation}
     -\left\langle i \omega_i^k \delta u^k_j \frac{\partial \delta u_i^k}{\partial r_j^k} \text{exp}\left\{ i \sum_{\alpha =1}^{N-1} \omega^{\alpha}_\mu \delta u^\alpha_\mu \right\} \right\rangle = i \frac{\partial^2 \psi}{\partial r_j^k \partial\omega^k_j}.
\end{equation}
The pressure and the vicious terms remain unclosed. The final form of the equation is
\begin{align}
    \frac{\partial \psi }{\partial t} - i \frac{\partial^2 \psi}{\partial r_j^k \partial\omega^k_j} = 
    \\
    i \sum_{k=1}^{N-1} \omega_i^k \Biggl \langle \Biggl\{ 
    -\left[ \frac{\partial p(\overline{x}^k)}{\partial x^k_i} -  \frac{\partial p(\overline{x}^0)}{\partial x^0_i}\right] + \nu \left( \frac{\partial^2}{\partial x_j^k \partial x_j^k} \delta u^k_i - \frac{\partial^2}{\partial x_j^0 \partial x_j^0} \delta u^0_i  \right)  \Biggl\} \text{exp}\left\{ i \sum_{\alpha =1}^{N-1} \omega^{\alpha}_\mu \delta u^\alpha_\mu \right\} \Biggr \rangle.
\end{align}

\section{Structure Function Equation from the Pressure Closure}
\label{appB_str_from_hopf_closure}

This appendix will derive the closed governing equation for the $2$-point structure function from the $2$-point velocity increment Hopf equation (\cref{eq:closed_2pt_hopf}) with the pressure closure given by \cref{eq:2pt_hopf_pressure_closure}. First apply the operator $\frac{\partial}{\partial \eta_3} \left( \frac{\partial}{\partial \eta_2} \right)^{2n-1} \eta_3$ to both sides of \cref{eq:closed_2pt_hopf} and then take the limits $\eta_2=\eta_3\rightarrow0$ and $\eta_1=r$. The operation applied to left hand side of \cref{eq:closed_2pt_hopf} is the same as in \cref{sec:SY21_restatement} and will yield the same terms on the left hand side of structure function equation from \cref{sec:SY21_restatement} (\cref{eq:closed_str_fun_eq}): focus instead on the terms generated by the pressure closure (\cref{eq:2pt_hopf_pressure_closure}). Notice that the characteristic function can be written as
\begin{equation}
    \psi = \sum_{k = 0}^\infty \left\langle \frac{\left(i \eta_2 \delta u\right)^k}{k!} e^{i \eta_3 \delta v} \right\rangle.
\end{equation}
Applying the operator to the first term of \cref{eq:2pt_hopf_pressure_closure} gives
\begin{equation}
    \frac{\partial}{\partial \eta_3} \left( \frac{\partial}{\partial \eta_2} \right)^{2n-1} \eta_3 a \frac{\partial}{\partial \eta_1} \frac{\partial}{\partial \eta_2} M_{s \rightarrow \eta_2}^{-1} \left[ \frac{2(s+1)}{s} M_{  \eta_2 \rightarrow s} \sum_{k = 0}^\infty \left\langle \frac{\left(i \eta_2 \delta u\right)^k}{k!} e^{i \eta_3 \delta v} \right\rangle \right].
\end{equation}
Simplifying gives:
\begin{equation}
    a \frac{\partial}{\partial \eta_1}\frac{\partial}{\partial \eta_3} \eta_3 \left( \frac{\partial}{\partial \eta_2} \right)^{2n} \left\langle e^{i \eta_3 \delta v} \sum_{k = 0}^\infty  \frac{\left(i \delta u \right)^k}{k!}  M_{s \rightarrow \eta_2}^{-1} \left[ \frac{2(s+1)}{s}  M_{  \eta_2 \rightarrow s}(\eta_2^k)   \right] \right\rangle.
\end{equation}
It can be shown that
\begin{equation}
    M_{  \eta_2 \rightarrow s}\eta_2^k = 2 \pi i \delta(s+k)
\end{equation}
and
\begin{equation}
     M_{s \rightarrow \eta_2}^{-1} \frac{-s-1}{-\frac{1}{2}s}2 \pi i \delta(s+k) = \frac{k-1}{\frac{1}{2}k} \eta_2^k.
\end{equation}
Therefore, the term equals:
\begin{equation}
    a \frac{\partial}{\partial \eta_1}\frac{\partial}{\partial \eta_3} \eta_3 \left( \frac{\partial}{\partial \eta_2} \right)^{2n} \left\langle e^{i \eta_3 \delta v} \sum_{k = 0}^\infty  \frac{\left(i \delta u \right)^k}{k!} \frac{k-1}{\frac{1}{2}k} \eta_2^k \right\rangle,
\end{equation}
Now the only nonzero term in $\left( \frac{\partial}{\partial \eta_2} \right)^{2n} \eta_2^k$ left over after taking $\eta_2 \rightarrow 0$ is $\frac{k!}{(k-2n)!}$ for $k=2n$. This leaves
\begin{equation}
    a \frac{\partial}{\partial \eta_1}\frac{\partial}{\partial \eta_3} \eta_3  \left\langle e^{i \eta_3 \delta v}   \frac{\left(i \delta u \right)^{2n}}{(2n)!} \frac{2n-1}{n} (2n)! \right\rangle,
\end{equation}
Applying the derivatives and then taking $\eta_3 \rightarrow 0$ and $\eta_1 =r$ give the result:
\begin{equation}
    a i^{2n} \frac{2n-1}{n}\frac{\partial S_{2n,0}}{\partial r}.
\end{equation}
The operator applied to the second term on the right hand side of \cref{eq:closed_2pt_hopf} is
\begin{equation}
    \frac{\partial}{\partial \eta_3} \left( \frac{\partial}{\partial \eta_2} \right)^{2n-1} \eta_3 \left( \frac{b}{r} \frac{\partial^2}{\partial \eta_3^2} \eta_2 \right)\psi,
\end{equation}
Applying the derivatives and taking the limits $\eta_2=\eta_3\rightarrow0$ and $\eta_1=r$, leaves the final expression:
\begin{equation}
    \frac{b}{r}(2n-1) i^{2n} S_{2n-2,2}.
\end{equation}
Putting it all together, including the additional structure function terms in \cref{eq:unclosed_str_fun_eq}, shows that applying the operator $\frac{\partial}{\partial \eta_3} \left( \frac{\partial}{\partial \eta_2} \right)^{2n-1} \eta_3$ to \cref{eq:closed_2pt_hopf} and then taking $\eta_2=\eta_3\rightarrow0$ and $\eta_1=r$ gives the structure function equation
\begin{align}
    \frac{\partial S_{2n,0}}{\partial r } + \frac{d-1}{r} S_{2n,0} - \frac{(d-1)(2n-1)}{r}S_{2n-2,2} =
    \\
    (2n-1)\frac{a}{n} \frac{\partial S_{2n,0}}{\partial r} - \frac{2n-1}{r} b S_{2n-2,2}.
\end{align}

\end{appen}\clearpage

\bibliographystyle{jfm}
\bibliography{jfm}



\end{document}